\documentclass[twocolumn,showpacs,english,prb,preprintnumbers,amsmath,amssymb,floatfix]{revtex4}
\usepackage[T1]{fontenc}
\usepackage[latin9]{inputenc}
\usepackage{babel}
\usepackage{graphicx}
\usepackage{epsfig,psfrag}
\usepackage{dcolumn}
\usepackage{bm}
\usepackage{color}
\usepackage{esint}

\begin{document}

\title{Supercurrent blockade in Josephson junctions with a Majorana wire}

\author{Alex Zazunov and Reinhold Egger}

\affiliation{ Institut f\"ur Theoretische Physik,
Heinrich-Heine-Universit\"at, D-40225  D\"usseldorf, Germany }

\date{\today}

\begin{abstract} We study the Josephson effect for a topologically nontrivial superconducting (TS) wire
with Majorana fermion end states and tunnel-coupled to $s$-wave BCS superconducting (S) electrodes.
In an S-TS junction, no supercurrent flow is possible under fairly general conditions.
For the S-TS-S junction, bulk TS quasiparticles must be accessible to have a Josephson effect.
In the noninteracting case, we derive the exact current-phase relation (CPR) and
find $\pi$-periodic behavior with negative critical current for 
weak tunnel couplings.
Charging effects then cause the anomalous CPR $I(\varphi)=I_c \cos\varphi$,
where the parity-sensitive critical current $I_c$ provides a signature for Majorana states.
\end{abstract}

\maketitle

\section{Introduction}

The exciting physics of topological insulator (TI) and topological superconductor (TS) materials currently enjoys a lot of attention.\cite{hasan,qi}  The zero-energy Majorana bound states (MBSs) located near the ends of a one-dimensional (1D)  TS wire have sparked immense theoretical activity,\cite{kitaev,beenakker} and first reports claiming experimental evidence for MBSs have appeared.\cite{sasaki}  Majorana fermions are special in that they are their own antiparticles: Majorana creation and annihilation operators are identical. MBSs may be useful for topological quantum computation\cite{nayak} and can induce spectacular nonlocal quantum phenomena.\cite{hasan,qi,kitaev,beenakker,fukane,fu,zazu}  Possible realizations of Majorana wires include semiconductor quantum wires with proximity-induced superconductivity for strong spin-orbit and Zeeman couplings,\cite{lutchyn,felix} nanowires made out of TIs (e.g., Bi$_2$Se$_3$) deposited on a superconductor,\cite{cook} and helical edge states in 2D hybrid HgTe/CdTe--superconductor quantum well structures.\cite{fu,fukaneqsh} A Majorana wire may effectively be realized also for a hole drilled into a TI film coated by a superconductor,\cite{feigelman} and for the vortex core state in a 2D noncentrosymmetric\cite{sato} or $p$-wave\cite{bolech} superconductor.  For a Majorana wire contacted by normal-conducting electrodes, the conductance exhibits resonant Andreev reflection when Coulomb charging effects are negligible,\cite{bolech,nilsson,law,flensberg,wimmer} electron teleportation under strong Coulomb blockade conditions,\cite{fu} and universal power law scaling in the intermediate regime.\cite{zazu} When two TS wires are contacted, one expects the fractional Josephson effect with $4\pi$-periodic current-phase relation (CPR), $I(\varphi)$, due to parity conservation.\cite{fukane,fukaneqsh,law2,vanheck} If the junction also contains a topologically trivial superconductor (TS-S-TS), additional periodicities may occur.\cite{jiang}

\begin{figure}[t]
\centering
\includegraphics[width=8cm]{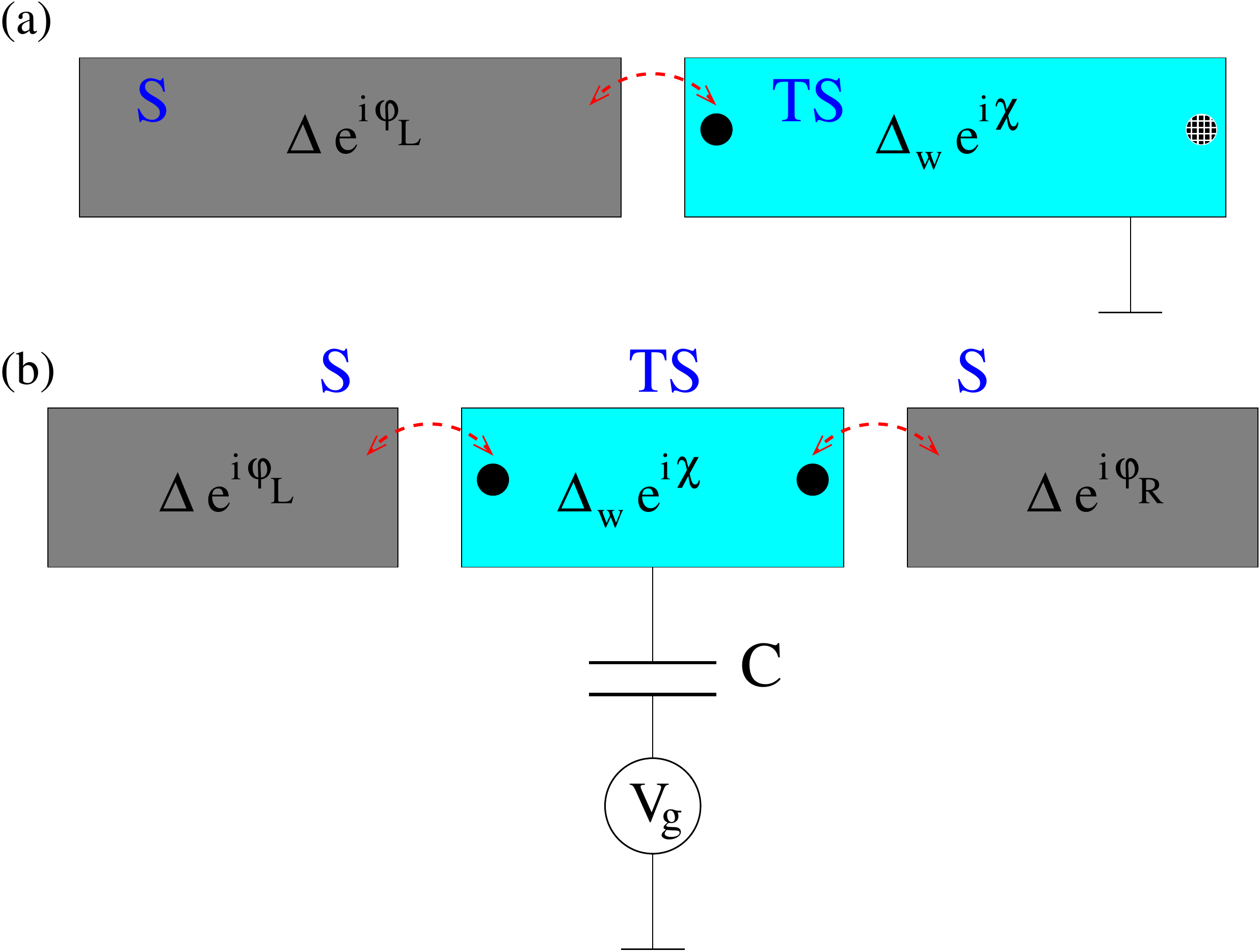}
\caption{\label{fig1} (Color online) Schematic Josephson junction setups involving a Majorana wire (TS) with proximity-induced gap $\Delta_w$ and phase $\chi$. Black dots stand for MBSs. The electrodes are $s$-wave BCS superconductors (S) with gap $\Delta$ and fixed phase $\varphi_{j=L/R}$.  Tunneling processes between an S electrode and the respective MBS and/or TS quasiparticle states are indicated by red dashed arrows. (a) S-TS junction. (b) S-TS-S junction with charging energy $E_c=e^2/(2C)$.  }
\end{figure}

In this work, we discuss the ground-state supercurrent flowing through the S-TS and S-TS-S junctions schematically shown in Fig.~\ref{fig1}. The left/right ($j=L/R$) electrodes correspond to standard $s$-wave BCS superconductors with (for simplicity identical) gap $\Delta$ and fixed phase $\varphi_{j}$.  The 1D Majorana wire contains a pair of decoupled MBSs at its ends. We consider a finite proximity-induced TS gap $\Delta_w$ and take into account the ($p$-wave type) TS quasiparticles.  For the S-TS-S junction with a floating (not grounded) Majorana wire, we also include a capacitive Coulomb interaction via the charging energy $E_c$. In practice, depending on the experimental realization, there can also be a parallel channel for Cooper pair  transfer involving only the superconducting substrate, and we here focus only on the Josephson current involving the TS wire.

Our main results are as follows.
(i) For the S-TS junction, the supercurrent is completely blocked for large $\Delta_w$ and/or for pointlike tunneling contacts. We provide a general condition explaining this supercurrent blockade. This has far-reaching consequences for the Josephson current through a Majorana wire whenever $s$-wave superconducting electrodes are involved.
(ii) In S-TS-S junctions, a finite supercurrent is only possible when quasiparticles (or other fermion excitations, e.g., a subgap impurity level) on the Majorana wire are accessible, at least for virtual transitions.
(iii) For the noninteracting ($E_c=0)$ S-TS-S junction, we provide the analytical solution for the CPR.  This solution shows that the fractional $4\pi$-periodic Josephson effect is absent in this setup. For weak tunnel couplings, we find a $\pi$-periodic CPR exhibiting $\pi$-junction behavior (negative critical current).
(iv) Expanding in the tunnel couplings and allowing for $E_c\ne 0$, we find the anomalous CPR
\begin{equation}\label{cosphi}
I(\varphi)=I_c \cos\varphi,
\end{equation}
with the critical current $I_c$ in Eq.~\eqref{critcur} below.  $I_c$
is parity-sensitive and could be used to detect Majorana fermion states.

The structure of this paper is as follows.  We first discuss the 
simplest S-TS junction in Sec.~\ref{sec2} and establish a 
general criterion for supercurrent blockade.  The S-TS-S junction
case is addressed in Sec.~\ref{sec3} by functional integral 
techniques. Besides allowing for some general conclusions,
this leads immediately to the exact Josephson CPR 
in the noninteracting limit, which we discuss in Sec.~\ref{sec4}.
In Sec.~\ref{sec5}, we then address the strongly interacting case by perturbation theory in the tunnel couplings.  We finally offer some conclusions
in Sec.~\ref{sec6}.  We mostly use units with $\hbar=1$ below.

\section{S-TS junction}\label{sec2}

We start with the grounded Majorana wire shown in Fig.~\ref{fig1}(a).  First, the BCS Hamiltonian describing the left S electrode is
\begin{equation}\label{bcs}
H_L = \sum_{\bf k} \Psi_{\bf k}^\dagger  \left( \xi_k \sigma_z+ 
\Delta\sigma_x e^{i\varphi_L\sigma_z } \right) \Psi_{\bf k},
\end{equation}
 with Nambu spinor $\Psi_{\bf k}= (c^{}_{{\bf k},\uparrow}, c^{\dagger}_{-{\bf k},\downarrow})^T$. Here, $c^\dagger_{{\bf k},\sigma}$ creates an electron with momentum ${\bf k}$, spin projection $\sigma=\uparrow,\downarrow=\pm$ and normal-state dispersion $\xi_k$. Pauli matrices $\sigma_{x,z}$ act in Nambu space.  

The TS wire contains a pair of decoupled zero-energy MBSs near its ends. These are described by Majorana fermion operators, 
$\gamma_{j=L/R}=\gamma_j^\dagger$, with anticommutator algebra 
\begin{equation}
\{\gamma_j,\gamma_{j'}\}= \delta_{jj'}.
\end{equation}
It is convenient to combine the two Majorana operators into a single auxiliary complex fermion, 
\begin{equation}
d=(\gamma_L+i\gamma_R)/\sqrt{2}. 
\end{equation}
We also include the bulk ($p$-wave type) TS quasiparticles.  The wire Hamiltonian is written in terms of right/left ($\alpha=r/l$) moving fermion operators $f^\dagger_{q,\alpha}$.  These describe TS quasiparticles with (discrete) 1D momentum $q$ and Fermi velocity $v_F$,
\begin{equation}\label{hw}
H_w=\sum_q \left( \begin{array}{c} f_{q,r}^\dagger \\ f_{-q,l}^{} \end{array} \right)^T
\left( \begin{array}{cc} v_F q &  \Delta_w e^{-i\chi} \\ \Delta_w e^{i\chi} & -v_F q  \end{array}\right)
\left( \begin{array}{c} f_{q,r}^{} \\ f_{-q,l}^{\dagger} \end{array} \right).
\end{equation}
Note that the zero modes $\gamma_{L,R}$ do not appear in $H_w$. 

Following, e.g., Ref.~\onlinecite{flensberg}, the S-TS contact is modeled by the  general tunneling Hamiltonian 
\begin{equation} 
H_t = \sum_{{\bf k}, \sigma \sigma'} \int dx \, c_{{\bf k},\sigma}^\dagger 
{\cal T}_{{\bf k},\sigma \sigma'}(x) \psi_{\sigma'}(x) + {\rm H.c.}
\end{equation}
with the field operator $\psi_\sigma(x)$ for electrons with 
spin projection $\sigma$ in the TS wire.  Note that the tunnel matrix elements encoded in the kernel ${\cal T}_{{\bf k},\sigma \sigma'}(x)$ may also describe spin-flip scattering at the interface.  Expanding $\psi_\sigma(x)$ in terms of the TS quasiparticle operators $f_{q,\alpha}$ and the Majorana fermions $\gamma_{j}$, we obtain
\begin{eqnarray} \label{ht}
H_t & = & \sum_{{\bf k}\sigma}
c_{{\bf k},\sigma}^\dagger \psi^{}_{{\bf k},\sigma} + {\rm H.c.},\\
\nonumber
\psi_{{\bf k},\sigma} & = & \lambda_{{\bf k},\sigma} \gamma_L +  \sum_{q,\alpha} t_{{\bf k},\sigma;q,\alpha} f_{q,\alpha}.
\end{eqnarray}
The complex-valued tunnel couplings $\lambda_{{\bf k},\sigma}$ 
and $t_{{\bf k},\sigma;q,\alpha}$ now encapsulate the overlap integrals  
of the kernel ${\cal T}_{{\bf k},\sigma\sigma'}(x)$ 
with the Majorana and quasiparticle wavefunctions in the TS wire,
respectively.  We stress that the spin properties of an arbitrary
TS wire (including the Majorana wavefunctions) as well
as spin-flip processes in the contact are fully taken into account via
the tunnel couplings $\lambda_{{\bf k}, \sigma}$ and $t_{{\bf k},\sigma;q,\alpha}$. 

Since $H=H_L+H_w+H_t$ corresponds to noninteracting fermions, the exact CPR, 
\begin{equation}\label{iphi}
I(\varphi)=(2e/\hbar) \partial_\varphi F(\varphi),
\end{equation}
can be obtained from the partition function, 
\begin{equation}\label{zphi1}
Z(\varphi)={\rm Tr}e^{-\beta H}=e^{-\beta F},
\end{equation}
with inverse temperature $\beta$ and phase difference 
$\varphi=\varphi_L-\chi$. After integration over the S electrons, 
the $\varphi$-dependent part of the free energy comes from the action piece
\begin{equation}\label{sts}
\beta{\cal F}_s = \int_0^\beta d\tau_1 d\tau_2 \sum_{\bf k} \Phi_{\bf k}^\dagger(\tau_1) G_k(\tau_1-\tau_2)\Phi_{\bf k}^{}(\tau_2) ,
\end{equation}
where $\Phi_{\bf k}=(\psi^{}_{{\bf k},\uparrow},\psi^\dagger_{-{\bf k},\downarrow})^T$ with Eq.~\eqref{ht}. The (Fourier-transformed) anomalous S Green's function is 
\begin{equation}
G_k(\omega) = \frac{\Delta}{\omega^2+\xi_k^2+\Delta^2} \sigma_x e^{i\varphi\sigma_z}.
\end{equation}
 Using $G_k(\tau)=G_k(-\tau)$ and $G_k\sim \sigma_{x,y}$, we find ${\cal F}_s=0$ whenever the time-reversed partners $\psi_{{\bf k},\uparrow}$ and $\psi_{-{\bf k},\downarrow}$ in Eq.~\eqref{ht} are collinear, 
\begin{equation}\label{collinear}
\psi_{-{\bf k},\downarrow}= \zeta_{\bf k} \psi_{{\bf k},\uparrow},
\end{equation} 
with some complex parameter $\zeta_{\bf k}$.\cite{foot1}  This is a sufficient (but not necessary) condition for supercurrent blockade in an S-TS junction.

To give some examples, consider the limit $\Delta_w\to \infty$, where no TS quasiparticles are accessible. In that case,  Eq.~\eqref{ht}
yields $\psi_{{\bf k},\sigma}=\lambda_{{\bf k},\sigma}\gamma_L$, which always  
satisfies Eq.~\eqref{collinear}.  In the absence of quasiparticles, an 
S-TS junction thus never carries a supercurrent.  Similarly, 
Eq.~\eqref{collinear} trivially holds for a spin-polarized TS wire,
where $\zeta_{\bf k}=0$.  However, for a point-like tunneling contact with
\begin{equation}\label{pointlike}
\lambda_{{\bf k},\sigma}\to \lambda_\sigma ,\quad 
t_{{\bf k},\sigma;\alpha,q}\to t_\sigma,
\end{equation} 
Eq.~\eqref{collinear} is generally not satisfied.  
Nonetheless, as we show in Sec.~\ref{sec4}, the supercurrent also 
vanishes in this limit. 

In order to rationalize why the supercurrent in a S-TS junction is blocked, 
we note that the condition (\ref{collinear}) implies 
a suppression of the Josephson effect unless noncollinear 
time-reversed states are available in the TS wire.  
However, the point-like tunneling example [Eq.~\eqref{pointlike}] also 
indicates that the supercurrent blockade occurs even when the noncollinearity 
requirement is met, and a full explanation of the general supercurrent
blockade is therefore more subtle.

\section{S-TS-S junction} \label{sec3}

In this section, we consider the Majorana wire between two S electrodes, see Fig.~\ref{fig1}(b), where $H_w$ is supplemented by the interaction term \cite{fu}
\begin{equation}\label{hd}
H_c = E_c\left(2\hat N+\hat n_d -N_0 + \sum_{q;i=1,2} \hat n^{(qp)}_{q,i}\right)^2.
\end{equation}
Cooper pairs in the wire are described by the number operator $\hat N$ with conjugate condensate phase $\chi$. Note that the operator $e^{-i\chi}$ ($e^{i\chi}$) lowers (raises) the Cooper pair number ($N$) by one unit.  The occupation of the $d$ fermion level corresponding to the two MBSs is given by $\hat n_d=d^\dagger d$, and the number $N_0$ in Eq.~\eqref{hd} can be continuously tuned by a backgate voltage, cf.~Fig.~\ref{fig1}(b).  The last term in Eq.~\eqref{hd} describes the occupation of TS quasiparticle states, $\hat n_{q,i}^{(qp)}=\eta_{q,i}^\dagger \eta_{q,i}^{}$, where Eq.~\eqref{hw} is written in diagonal form, 
\begin{equation}
H_w = \sum_{q}\sum_{i=1,2} E_q \eta^\dagger_{q,i}\eta_{q,i}^{},
\end{equation}
with fermion operators $\eta_{q,1}$ (particle-like) and $\eta_{q,2}$ (hole-like) for positive energy 
\begin{equation}\label{Eqdef}
E_q=\sqrt{v_F^2 q^2 + \Delta_w^2}.
\end{equation}
The $\eta_{q,i}$ are connected to $f_{q,\alpha}$ by a 
canonical charge-conserving transformation,
\begin{equation}\label{connect}
\left( \begin{array}{c} f_{q,r} \\ e^{-i\chi} f^\dagger_{-q,l} \end{array}\right) = \left( \begin{array}{cc} u_q & -v_{q} \\ v_q & u_q\end{array}\right)
\left(\begin{array}{c} \eta_{q,1}\\ e^{-i\chi} \eta_{q,2}^\dagger \end{array}\right),
\end{equation}
where 
\begin{equation}
u_q=v_{-q}=\sqrt{(1+v_F q/E_q)/2}.
\end{equation} 
The full Hamiltonian, 
\begin{equation}
H=H_L+H_R+H_w+H_c+H_t ,
\end{equation}
also includes a BCS term $H_R$ for the right S electrode, see Eq.~\eqref{bcs}, with
 $\varphi\equiv \varphi_L-\varphi_R$, and the tunneling Hamiltonian 
\begin{equation}
H_t=H_t^{(\gamma)}+H_t^{(qp)}.
\end{equation}
In what follows, point-like tunneling contacts satisfying Eq.~\eqref{pointlike} are assumed for simplicity. Taking into account charge conservation,\cite{zazu}
\begin{eqnarray}\label{htg}
H_t^{(\gamma)}&=& \frac{1}{\sqrt{2}} \sum_{\bf k} \Bigl [ \lambda_{L,\sigma} c^\dagger_{{\bf k},L,\sigma}
\left ( d + e^{-i\chi} d^\dagger\right) \\ \nonumber &-& i\lambda_{R,\sigma} c_{{\bf k},R,\sigma}^\dagger
\left ( d - e^{-i\chi} d^\dagger\right) \Bigr] + {\rm H.c.}
\end{eqnarray}
describes tunneling between states in lead $j=L/R$ and the corresponding Majorana fermion $\gamma_j$ with amplitude $\lambda_{j,\sigma}$.  Quasiparticle tunneling is contained in [see Eqs.~\eqref{ht} and \eqref{connect}] 
\begin{eqnarray}\label{qptune}
H_t^{(qp)} &=& \sum_{{\bf k},j,\sigma} t_{j,\sigma} c^\dagger_{{\bf k},j,\sigma} \eta_w + {\rm H.c.},\\ \nonumber
\eta_w &=&  \sum_q \left[ u_q \left( \eta_{q,1}+\eta_{q,2}\right)+v_q 
e^{-i\chi}\left(\eta^\dagger_{q,1}- \eta^\dagger_{q,2}\right)\right].
\end{eqnarray}
With the (normal-conducting) lead density of states $\nu_0$, we introduce the hybridization scales
\begin{eqnarray}\nonumber
\Gamma_j & = & 2\pi\nu_0 \sum_\sigma \sigma \lambda_{j,\sigma} t_{j,-\sigma}^*,
\quad \tilde\Gamma_j = 2\pi \nu_0\sum_\sigma \lambda_{j,\sigma} t_{j,\sigma}^*,\\
\label{hyb} \Gamma_{\lambda,j} &=& 2\pi\nu_0\sum_\sigma |\lambda_{j,\sigma}|^2,\quad
\Gamma_t = 2\pi\nu_0\sum_{j,\sigma} |t_{j,\sigma}|^2.
\end{eqnarray}

Now we are ready to address the CPR. For $E_c\ne 0$, no exact solution can be obtained anymore, but general insights follow again by considering the partition function $Z(\varphi)$ in Eq.~\eqref{zphi1}. Moreover, the exact CPR for $E_c=0$ reported in Sec.~\ref{sec4} directly follows from our expressions below.  After integration over the lead fermions,
\begin{equation}\label{zphi}
Z(\varphi)={\rm Tr}\left( e^{-\beta(H_w+H_c)} {\cal T} e^{-(S_M+S_{qp}+S_{\rm int})} \right),
\end{equation}
where ${\cal T}$ denotes imaginary-time ($\tau$) ordering and the trace indicates functional integration over the phase field $\chi(\tau)$, Majorana fields $\gamma_{L/R}(\tau)$, and quasiparticle fields $(\eta_w,\bar \eta_w)(\tau)$. Using the hybridization parameters \eqref{hyb} and the auxiliary function 
\begin{equation}\label{ftau}
f(\tau)= \int \frac{d\omega}{2\pi} f_\omega e^{-i\omega\tau},\quad
f_\omega = \frac{1}{\sqrt{\omega^2+\Delta^2}},
\end{equation}
the action $S_M$ in Eq.~\eqref{zphi} comes from the Majorana fields only,
\begin{eqnarray} \label{smm}
S_M &= & \int d\tau_1 d\tau_2 \dot f(\tau_1-\tau_2) \\ \nonumber
&\times& \sum_{j,j'} 
\Lambda_{j,j'}(\chi_1,\chi_2) \gamma_j(\tau_1) \gamma_{j'}(\tau_2),
\end{eqnarray}
where $j=L/R=+/-$ and $\chi_{i=1,2}=\chi(\tau_{i})$. 
The matrix kernel $\Lambda$ is given by 
\begin{eqnarray*}
\Lambda_{j,j} &=& \sum_{j'}\frac{\Gamma_{\lambda,j'}}{8} 
[ 1+\cos(\delta\chi)+\sigma_{jj'} (\cos\chi_1+\cos\chi_2)],\\
\Lambda_{j,-j}&=& \pm \sum_{j'} \frac{\Gamma_{\lambda,j'}}{8} 
[\sin(\delta\chi)-\sigma_{jj'} (\sin\chi_1+\sin\chi_2)],
\end{eqnarray*}
where $\sigma_{jj'}={\rm sgn}(jj')$ and $\delta\chi=\chi_1-\chi_2$.
Importantly, $S_M$ in Eq.~\eqref{smm} does not depend on the phase difference $\varphi$.  For $\Delta_w\to \infty$, the action piece $S_{qp}+S_{\rm int}$ does not contribute, and therefore there is no Josephson effect in this limit. 

Turning to finite $\Delta_w$, $H_t^{(qp)}$ results in 
\begin{equation}
S_{qp}=\Gamma_t \int d\tau_1 d\tau_2 \dot f(\tau_1-\tau_2) \bar\eta_w(\tau_1)\eta_w(\tau_2),
\end{equation} 
which is also independent of $\varphi$.  The only $\varphi$-dependent action piece, which ultimately can be responsible for a Josephson current, involves Majoranas and quasiparticles,
\begin{equation}\label{sint1}
S_{\rm int} = \int \frac{ d\tau_1 d\tau_2 }{2\sqrt{2}} 
\sum_{j} \gamma_j(\tau_1) L_{j}(\tau_1,\tau_2)\eta_w(\tau_2) + {\rm H.c.},
\end{equation}
with the kernels ($j=L/R=+/-$)
\begin{eqnarray} \label{ljfunc}
L_j(\tau_1,\tau_2) &=& \tilde\Gamma_j^* \dot f(\tau_1-\tau_2) \\ \nonumber
&\times& \left[ 1+e^{-i\chi_1} \pm i(1-e^{-i\chi_1}) \right]\\ \nonumber
&+&  i \Delta\Gamma_j^* f(\tau_1-\tau_2)\Bigl [ ie^{i\varphi_j}(1+e^{i\chi_1}) 
\\ \nonumber && \pm e^{i\varphi_{-j}}(1-e^{i\chi_1}) \Bigr ].
\end{eqnarray}
This implies that the magnitude of the Josephson current is controlled by the ratio $|\Gamma_{L}\Gamma_R|/\Delta^2_w$. In particular, $I=0$ when this ratio vanishes.  Equation (\ref{ljfunc}) also tells us how $s$-wave Cooper pairs are transported through the Majorana wire, namely by creation ($d^\dagger \eta_w^\dagger$) or annihilation ($e^{-i\chi}d \eta^\dagger_w$) of charge $2e$ (plus the conjugate processes).  These processes simultaneously involve Majorana states and bulk TS quasiparticles.

\section{Noninteracting limit}\label{sec4}

For $E_c=0$, we again have a system of noninteracting fermions. The exact $T=0$ CPR follows after some algebra from the expressions in Sec.~\ref{sec3}, 
\begin{equation}\label{nint}
I(\varphi) = -\frac{2e}{\hbar} \int_0^\infty \frac{d\omega}{2\pi} \partial_\varphi \ln \det M(\omega) ,
\end{equation}
with the matrix $M_{jj'}(\omega)$ in left/right ($j=L/R=+/-$) space. 
Using the notation [see also Eqs.~\eqref{Eqdef}, \eqref{hyb} and \eqref{ftau}]
\begin{equation}
X_\omega =  \left( \sum_q \frac{1}{\omega^2+E_q^2} \right)^{-1} + 
\omega^2 f_\omega \Gamma_t, 
\end{equation}
the diagonal elements of $M$ are given by 
\begin{equation}
M_{jj} = 1+f_\omega \Gamma_{\lambda,j} + \frac{f_\omega^2}{X_\omega}
\left(\Delta^2\left|\Gamma_j\right|^2-
\omega^2 \left|\tilde\Gamma_j\right|^2\right).
\end{equation}
Similarly, we find for the off-diagonal elements
\begin{eqnarray}\label{offdiagM}
M_{j,-j} & = & \frac{f_\omega^2}{X_\omega} \ 
{\rm Re}\Bigl [ \Delta^2\Gamma_j^* \Gamma_{-j} e^{ \pm i\varphi} 
-\omega^2\tilde\Gamma_j^*\tilde\Gamma_{-j} \\ \nonumber
&+& i\omega \Delta\left(\Gamma_j^*\tilde\Gamma_{-j} e^{i \varphi_j} 
+\tilde\Gamma_j^*\Gamma_{-j}e^{- i\varphi_{-j}} \right)\Bigr].
\end{eqnarray}
We now draw several conclusions from Eq.~\eqref{nint}.  

First, for our point-like tunneling model, since only the off-diagonal matrix 
elements [Eq.~\eqref{offdiagM}] depend on the phase difference $\varphi$, 
a single S-TS junction, where the Majorana fermion $\gamma_R$ is effectively
not accessible, cannot carry a supercurrent.  This result was already
announced in Sec.~\ref{sec2}.

Second, exploiting current conservation, some algebra reveals the periodicity 
\begin{equation}
I(\varphi + n \pi) = I(\varphi)
\end{equation} 
in Eq.~\eqref{nint}, with $n = 2$ or just 1 (see below).  
In particular, the S-TS-S junction
does not allow for the fractional $4 \pi$-periodic Josephson effect 
omnipresent in TS-TS junctions.\cite{beenakker}

Third, for a symmetric junction with $\Gamma_L=\Gamma_R=\Gamma$, 
expanding Eq.~\eqref{nint} to lowest nontrivial order
in the tunnel couplings yields\cite{foot2}
\begin{equation}\label{grcprec0}
I(\varphi) = - \frac{e\Delta}{\hbar} 
\left|\pi \nu_w \Gamma^2/\Delta_w \right|^2 
{\cal G}(\Delta_w/\Delta) \sin(2\varphi)
\end{equation}
 with the density of states $\nu_w = L/(2 \pi v_F)$ in a wire of length $L$ 
and the function
\begin{equation}
{\cal G}(x)=\frac{ x (2+x)}{4(1+x)^2} .
\end{equation}
Equation (\ref{grcprec0}) describes a $\pi$-periodic CPR with negative critical current, $I_c \sim - |\Gamma/\Delta_w|^4.$  It is worth mentioning that $I_c$ is suppressed by the factor $|\Gamma/\Delta_w|^2$ compared
to the usual cotunneling limit. 
This suppression can be traced back to the destructive interference of different contributions of order $\Gamma^2$.
For finite $E_c$, this cancellation is incomplete and we recover $I_c \sim |\Gamma/\Delta_w|^2$, see Eq.~\eqref{critcur}.

\section{Weak tunneling limit}\label{sec5}

In this section, we discuss the results of second-order perturbation theory in the hybridizations \eqref{hyb}, valid for arbitrary charging energy $E_c$.  In the ground state, no quasiparticles are thermally activated but virtual excitations remain possible.  For simplicity, we here only keep $q=0$ quasiparticles\cite{foot3} and consider the case $\Gamma_L=\Gamma_R=\Gamma$. The ground-state energy of the isolated dot can then be expressed in terms of the energies 
\begin{equation}
E_n=E_c(2N-N_0+n)^2 , \quad n\in\mathbb{Z},
\end{equation} 
where the Cooper pair number $N$ follows from the relation $-3/2<2N-N_0<1/2$.
Using $\delta\equiv 2N-N_0$, for $-1/2<\delta<1/2$, the ground state $|e\rangle$ has energy $E_0$ and even parity, i.e., the $d$ state is empty.  Otherwise, we have an odd-parity state, $|o\rangle=d^\dagger|e\rangle$, with energy $E_1$.  

\begin{figure}[t]
\centering
\includegraphics[width=8cm]{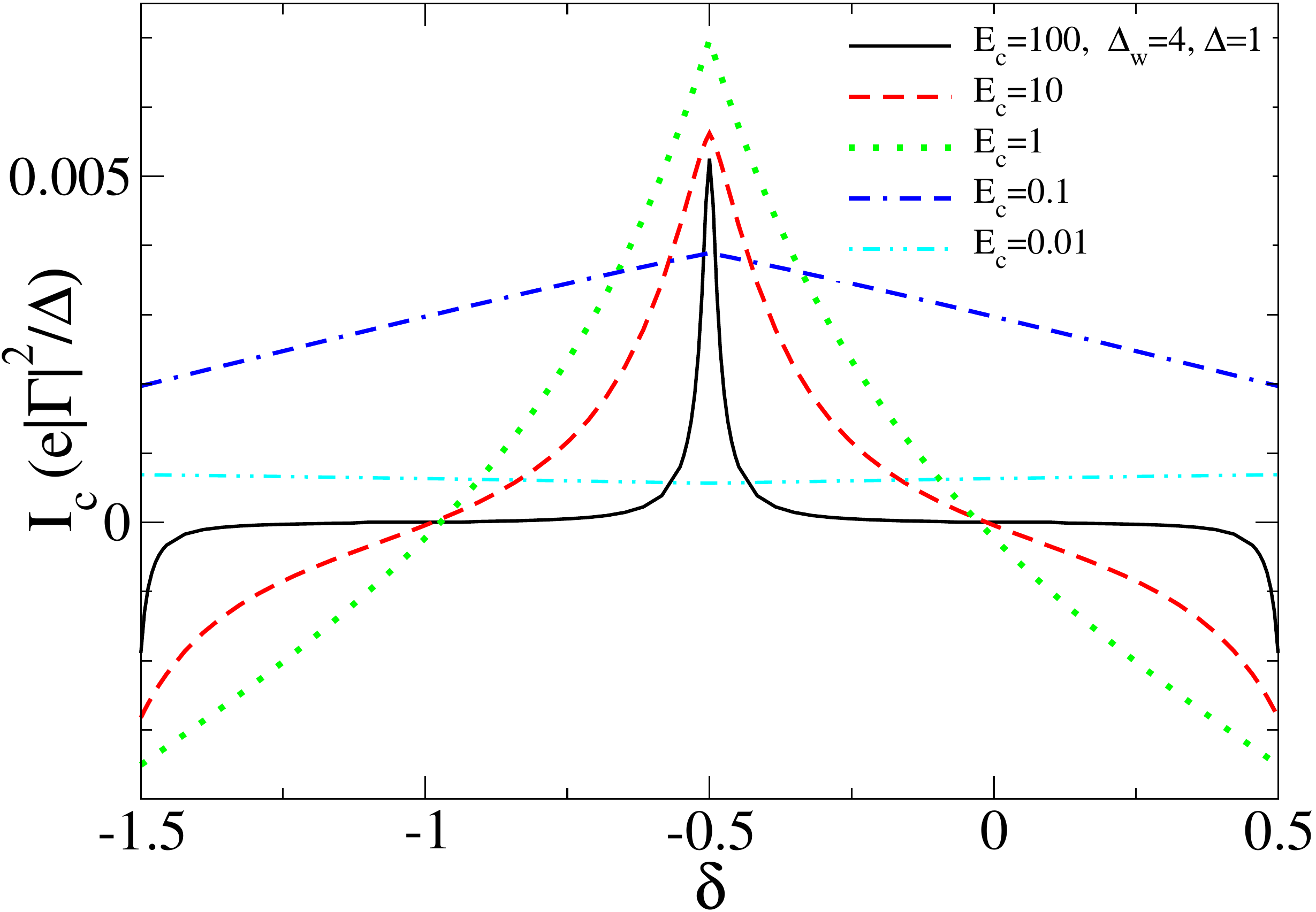}
\caption{ \label{fig2} (Color online) Anomalous current $I_c$ in
Eq.~\eqref{cosphi} vs $\delta = 2N-N_0$ in the cotunneling regime of
an S-TS-S junction. $I_c$ is computed from Eq.~\eqref{critcur} for
$\Delta_w/\Delta=4$ and several $E_c/\Delta$.  Note that $\delta$
can be changed by varying a backgate, see Fig.~\ref{fig1}(b).
$I_c(\delta)$ is periodic; we show one fundamental interval only.}
\end{figure}
\begin{table}[t]
\begin{tabular}{|c|c|c|}
\hline $Q_1$ & $Q_2$ & $Q_3$ \\ \hline \hline
$+ \left( \hat \epsilon_1, \hat \epsilon_2, \hat \epsilon_1 \right)$ &
$- \left( \epsilon_1, 0, \hat \epsilon_{-1} \right)$ &
$+ \left( \epsilon_1, 0, \hat \epsilon_1 \right)$ \\
$- \left( \hat \epsilon_{-1}, \hat \epsilon_2, \hat \epsilon_{-1} \right)$ &
$+ \left( \epsilon_3, 0, \hat \epsilon_1 \right)$ &
$- \left( \epsilon_3, 0, \hat \epsilon_{-1} \right)$ \\
$+ \left( \epsilon_1, \hat \epsilon_2, \epsilon_1 \right)$ &
$- \left( \hat \epsilon_{-1}, 0, \epsilon_1 \right)$ &
$+ \left( \hat \epsilon_1, 0, \epsilon_1 \right)$ \\
$- \left( \epsilon_3, \hat \epsilon_2, \epsilon_3 \right)$ &
$+ \left( \hat \epsilon_1, 0, \epsilon_3 \right)$ &
$- \left( \hat \epsilon_{-1}, 0, \epsilon_3 \right)$ \\
$- \left( \epsilon_1, \hat \epsilon_2, \hat \epsilon_1 \right)$ &
$+ \left( \epsilon_1, \Delta_w, \hat \epsilon_{-1} \right)$ &
$- \left( \hat \epsilon_{-1}, \Delta_w, \hat \epsilon_{-1} \right)$ \\
$+ \left( \epsilon_3, \hat \epsilon_2, \hat \epsilon_{-1} \right)$ &
$- \left( \epsilon_3, \hat \epsilon_4, \hat \epsilon_1 \right)$ &
$+ \left( \hat \epsilon_1, \hat \epsilon_4, \hat \epsilon_1 \right)$ \\
$- \left( \hat \epsilon_1, \hat \epsilon_2, \epsilon_1 \right)$ &
$+ \left( \hat \epsilon_{-1}, \Delta_w, \epsilon_1 \right)$ &
$- \left( \epsilon_1, \Delta_w, \epsilon_1 \right)$ \\
$+ \left( \hat \epsilon_{-1}, \hat \epsilon_2, \epsilon_3 \right)$ &
$- \left( \hat \epsilon_1, \hat \epsilon_4, \epsilon_3 \right)$ &
$+ \left( \epsilon_3, \hat \epsilon_4, \epsilon_3 \right)$ \\
\hline
\end{tabular}
\caption{ \label{table1} All combinations of signs $\sigma_p = \pm$ and energy arguments $(E_a,E_b,E_c)$ summed over in Eq.~\eqref{critcur} for the even-parity state, where $\hat \epsilon_n= \epsilon_n+\Delta_w$ and $\epsilon_n=E_n-E_0$.  }
\end{table}

Some straightforward but lengthy algebra yields from Eq.~\eqref{zphi} the CPR announced in Eq.~\eqref{cosphi}.  For the even-parity state $|e\rangle$,  the critical current is
\begin{equation}\label{critcur}
I_c= \frac{e\Delta}{\hbar} \ \frac{\Delta |\Gamma|^2}{2}
\sum_{p=1}^{24} \sigma_p Q_{i_p} \left(E_{a_p},E_{b_p}, E_{c_p}\right).
\end{equation}
All signs $\sigma_p=\pm$ and energies $(E_a,E_b,E_c)$ entering the real-valued $Q$ functions are given in Table \ref{table1}, where
\[
Q_{i_p}= \int\frac{d\omega d\omega'}{(2\pi)^2} \frac{f_\omega f_{\omega'}} {i\omega+E_a} \times
\left\{ \begin{array}{cc}
\frac{1-\delta_{E_b,0}}{E_b (i\omega'+E_c)},& i_p=1,\\
\frac{1}{(i\omega+E_c)(i\omega+i\omega'+E_b)},& i_p=2, \\
\frac{1}{(i\omega'+E_c)(i\omega+i\omega'+E_b)},& i_p=3.
\end{array} \right .
\]
For the odd-parity state $|o\rangle$, $I_c$ follows from Eq.~\eqref{critcur} with a particle-hole transformation, $\epsilon_n\to \epsilon_{-(n-1)}$, in Table \ref{table1}.

The critical current $I_c$ in Eq.~\eqref{critcur} is shown in Fig.~\ref{fig2} as a function of $\delta$. For half-integer $\delta$, two charge states become degenerate and $|I_c|$ shows resonance enhancement.  While for small $E_c$, we find a small positive and $\delta$-independent $I_c$, these resonances become very narrow for large $E_c$, with $I_c$ close to zero unless $\delta$ is nearly half-integer. The parity-sensitivity of $I_c$, i.e., the sign change
of $I_c$ between $\delta\approx -1/2$ and $\delta\approx 1/2$ (mod 2), see
Fig.~\ref{fig2}, may then provide an experimentally 
detectable signature for MBSs.

\section{Conclusions}\label{sec6}

In this work, we have shown that under generic conditions, a supercurrent blockade occurs in S-TS Josephson junctions made by coupling an $s$-wave BCS superconductor and a topologically nontrivial superconducting wire with Majorana end states. In the S-TS-S junction, this blockade imposes severe restrictions; in particular, quasiparticle excitations are necessary to have a 
Josephson effect.  For the noninteracting case, the analytical solution for
the current-phase relation has been presented. It shows that 
the $4\pi$-periodic Josephson effect is absent in S-TS-S junctions.
For weak tunneling, the CPR is $\pi$-periodic with a negative critical current, while a finite charging energy results in
the anomalous CPR  $I(\varphi)=I_c\cos\varphi$. The parity sensitivity of 
$I_c$ could then be used to detect the Majorana states in supercurrent 
measurements.  We hope that our predictions will soon be tested experimentally.

\acknowledgments

We thank B. Braunecker, K. Flensberg and A. Levy Yeyati for helpful discussions.
This work was supported by the DFG (Grant No. EG 96/9-1 and SFB TR 12).


\begin{thebibliography}{12}
\bibitem{foot2}
For an asymmetric junction, 
one generally finds a shifted CPR. This can be seen 
by expanding to lowest nonvanishing order in 
$\Gamma_j$ and $\tilde \Gamma_j$.   We  then find
$I(\varphi) = | \Gamma_L \Gamma_R|^2 F(2 \varphi) +
\sum_j | \tilde \Gamma_j \Gamma_{-j}|^2 F_j(\varphi)$, with the
$\pi$-periodic function $F(2 \varphi) \sim\sin (2 \varphi + 2 \alpha)$,
where $\alpha = {\rm arg} \left( \Gamma^\ast_L \Gamma_R \right)$,
and $2\pi$-periodic functions $F_{L/R}(\varphi)$.
\bibitem{foot3}
It is straightforward to include $q\ne 0$ quasiparticles, but their effect is weak unless both $E_c$ and $\Delta_w$ are small.
\end{thebibliography}
\end{document}